\documentstyle[aps,prl,epsf]{revtex}

\newcommand{\be}{\begin{equation}}
\newcommand{\ee}{\end{equation}}
\newcommand{\ba}{\begin{eqnarray}}
\newcommand{\ea}{\end{eqnarray}}

\begin{document}

\draft

\twocolumn[\hsize\textwidth\columnwidth\hsize\csname@twocolumnfalse%
\endcsname

\title{Comment on ``Effects of Disorder on Ferromagnetism in Diluted
Magnetic Semiconductors''}
\author{C. Timm, F. Sch\"afer, and F. von Oppen}
\address{Institut f\"ur Theoretische Physik, Freie Universit\"at
Berlin, Arnimallee 14, D-14195 Berlin, Germany}
\date{September 5, 2001}

\maketitle

\begin{abstract}
\end{abstract}
\pacs{75.50.Pp,71.30.+h,75.40.Mg}

]

\narrowtext

In a recent Letter, Berciu and Bhatt \cite{BB} have presented a mean-field
theory of ferromagnetism in III-V semiconductors doped with Mn \cite{Ohno}.
The approach starts from uncorrelated Mn impurity states, which overlap to
form an impurity band (IB). We argue here that this approach is not
appropriate for (Ga,Mn)As containing 1--5\% Mn. There is also a sign error
in Ref.~\cite{BB}. After correcting this error, mean-field theory no longer
yields stable ferromagnetism for the system sizes used.

\begin{figure}[h]
\centerline{\epsfxsize 5in\epsfbox{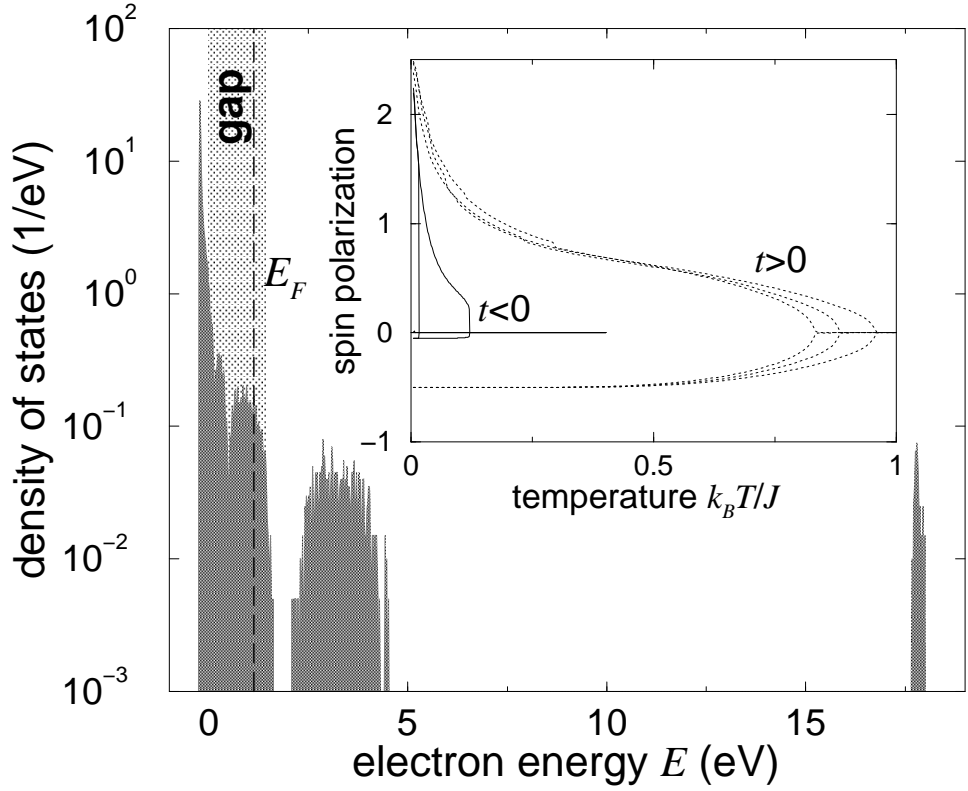}}
\caption{Density of states per Mn impurity as a function of {\it
electron\/} energy relative to the valence band top for $x=0.05$,  averaged
over $100$ configurations of $200$ Mn impurities (dark gray area). The
light gray bar denotes the GaAs gap. The dashed line shows the
Fermi energy for $p=0.1$. Inset: Average Mn (positive) and
hole (negative) spin polarizations for the same parameters. The dotted
curves have been obtained with overlap matrix elements $t(r)>0$ as in
Ref.~\protect\cite{BB}, the solid curves with the correct $t(r)<0$, each
for three configurations.}
\label{fig.1}
\end{figure}

For the overlap matrix elements between hole impurity states,
Ref.~\cite{BB} gives $t(r)=2E_b\,(1+r/a_B)\,e^{-r/a_B}$ with hole binding
energy $E_b$ and Bohr radius $a_B$. The calculations are done with
$t(r)>0$. We have reproduced the numerical results of Ref.~\cite{BB}, for
example, the dotted curves in the inset in Fig.~\ref{fig.1} give the Mn and
hole spin polarizations for three impurity configurations with a Mn
concentration of $x=0.05$ and $p=0.1$ holes per Mn, using $200$ Mn spins.
These are evidently the same parameters as used for Fig.~3 of
Ref.~\cite{BB}.

However, the overlap matrix elements $t(r)$ should be negative, as we show
now, using the sign convention of Ref.~\cite{BB}. For a parabolic valence
band, the Hamiltonian for a hole and two impurities at ${\bf R}_A$ and
${\bf R}_B$ reads $H = p^2/2m^\ast + V_A({\bf r}) + V_B({\bf r})$ with
$V_i({\bf r})=-e^2/\epsilon|{\bf r}-{\bf R}_i|$ where $i=A,B$. $m^\ast$ is
the effective mass and $\epsilon$ is the dielectric constant. The
hydrogen-like Hamiltonian for a single impurity is $H_i = p^2/2m^\ast +
V_i({\bf r})$. Let $\psi_i$ be the ground state of $H_i$ with energy
$E_b<0$. The overlap matrix element between $\psi_A$ and $\psi_B$ is
$t\equiv \langle \psi_B|H|\psi_A\rangle$, which we can also write as $t =
\langle\psi_B|\, H_A + V_B({\bf r}) |\psi_A\rangle =
E_b\langle\psi_B|\psi_A\rangle - \langle\psi_B| e^2/\epsilon|{\bf r}-{\bf
R}_B|\,|\psi_A\rangle$. Since the wave functions $\psi_i$ are real and
positive, we obtain $t<0$.

%Also, the prefactor in the
%expression for $t(r)$ in Ref.~\cite{BB} should be $3$ instead of $2$
%\cite{Bhatt}.

The sign change in $t(r)$ inverts the {\it highly asymmetric\/} IB. With
the correct sign the IB has a long tail to negative hole energies (positive
electron energies), resulting from strongly bound states. Fig.~\ref{fig.1}
shows the density of states (DOS) for $x=0.05$ with $200$ Mn impurities.
The subband structure is due to the finite system size. We find that the IB
width (independent of the sign of $t$) is about $18\:\mathrm{eV}$, and thus
much larger than the gap of GaAs. Furthermore, the IB strongly overlaps
with the valence band. These results show that the IB picture is not
appropriate. The reason is that there are typically many Mn impurities
within the range over which $t(r)$ falls off, {\it i.e.}, the system is
{\it not\/} weakly doped \cite{SE}.

%A more realistic approach should
%probably describe the random Coulomb potential due to the charged
%impurities and then consider valence-band holes in this potential.

With the corrected $t(r)$ the Fermi energy no longer lies in a region of
very large DOS as in Ref.~\cite{BB}, but in the band tail close to the
conduction band edge, as shown in
Fig.~\ref{fig.1} for $p=0.1$. Average Mn and hole spin polarizations for
$x=0.05$ and $p=0.1$ are shown in the inset. The maximum Curie temperature
is only about $20\:\mathrm{K}$ and the stability of the ferromagnetic state
strongly depends on the impurity configuration. If ferromagnetism is found
at all, the hole polarization is at most $0.05$. For the numerical parameters
this means that only a {\it single\/} hole spin is flipped, suggesting large
finite-size effects.

To conclude, the impurity-band approach is not applicable to (Ga,Mn)As
containing a few percent of Mn. However, it should be valuable for weak
doping. We would like to thank J. K\"onig, M. Raikh, and J. Schliemann for
interesting discussions.

%%% FIGURES

\newpage
\widetext

\end{document}